\documentclass[superscriptaddress,showpacs,aps,pra]{revtex4}
\usepackage{amsmath}
\usepackage{graphicx}
\usepackage[utf8]{inputenc}
\usepackage{bm}
\usepackage{comment}
\begin{document}
\title{The Casimir Effect for Arbitrary Optically
  Anisotropic Materials}  
\date{\today}
\author{José C. Torres-Guzmán}
\email{torres@fis.unam.mx}
\affiliation{Instituto de Ciencias Físicas, Universidad Nacional
   Autónoma de México,\\Apdo. Postal 48-3, 62251 Cuernavaca,
   Morelos, México}
\affiliation{Facultad de Ciencias, Universidad Aut\'onoma del Estado
de Morelos, Avenida Universidad 1001, 62221 Cuernavaca, Morelos, M\'exico}
\author{W. Luis Mochán}
\email{mochan@fis.unam.mx}
\affiliation{Instituto de Ciencias Físicas, Universidad Nacional
   Autónoma de México,\\Apdo. Postal 48-3, 62251 Cuernavaca,
   Morelos, México}

\begin{abstract}
We extend a fictitious-cavity approach to calculate the Casimir effect
for cavities bounded by flat anisotropic materials. We calculate the 
energy,
force and torque in terms only of
the optical coefficients of the walls of the cavity. 
We calculate the Casimir effect at zero and finite
temperature for some simple systems. As a non trivial
application, we calculate the torque between a semi-infinite
anisotropic plate and an anisotropic film. We study the
effect of the film thickness in the torque and find an optimal width
that maximizes the torque. 
\end{abstract}
%Ellos sugirieron
% Nuestros originales \pacs{42.50.Lc,12.20.Ds,42.25.Lc,42.50.Tx}
%*42.50.Pq  %Cavity quantum electrodynamics; micromasers 
%*31.30.jh  %QED corrections to long-range and weak interactions
%**42.50.Lc, %quantum fluctuations
%12.20.Ds, % quantum electrodynamics, specific calculations
%42.25.Lc, % 	Birefringence
%42.50.Tx %	Optical angular momentum and its quantum aspects (see also
%12.20.-m, %	Quantum electrodynamics
%78.68.+m,  % 	Optical properties of surfaces
%42.50.Nn  % 	Quantum optical phenomena in absorbing...
%07.10.Pz, % 	Instruments for strain, force, and torque
%46.55.+d, % 	Tribology and mechanical contacts
%03.50.De,% 	Classical electromagnetism
%03.65.Sq,% 	Semiclassical theories and applications
%42.50.Pq,% 	Cavity quantum electrodynamics; micromasers
%73.20.Mf,% 	Collective excitations...

\maketitle

\section{Introduction}
An electromagnetic mode of frequency $\omega$ within an
electromagnetic cavity is analogous to a harmonic oscillator with a
quantized energy spectrum given by semi-integer multiples $n+1/2$ of the
energy quantum $\hbar \omega$, where the integer $n$ is the occupation
number of the mode, i.e., the number of photons in the corresponding
state, $\hbar \omega$ is the quantized energy of each photon
and $\hbar\omega/2$ is the ground state energy arising from the
quantum nature of the electromagnetic field and its {\em zero
point fluctuations}. As the frequency of the electromagnetic modes
depends on the geometry of the cavity, the zero point fluctuations may
not be simply disregarded as a constant contribution to the energy 
and should be accounted for in calculations of the total energy of the
system. A simple vacuum cavity may be produced by positioning two flat
conducting plates parallel to each other a small distance $L$ apart,
leading to a first quantization of the electromagnetic field within
the cavity.  In 1948, Casimir \cite{casimir} predicted that the vacuum 
electromagnetic energy due to the quantum fluctuations of the field 
would depend on $L$ and therefore a force, which turns out to be
attractive, would act on each plate.  This Casimir force also may
be conceived as originated from the difference between the radiation pressure
due to the fluctuating electromagnetic modes outside the plates and
the modes within \cite{milloni}. Irrespective of its interpretation,
the Casimir force 
has its origin in the linear momentum carried by the radiation
field. Nevertheless, the 
radiation field carries angular momentum beyond linear
momentum. Therefore, if the plates 
are optically anisotropic, a Casimir torque might develop besides the
Casimir force.  In 
fact, the transfer of angular momentum of polarized light to a macroscopic
birefringent medium, resulting in a torque, has been known for a long
time \cite{beth}. The Casimir torque may be interpreted as arising from the
orientational dependence of the vacuum electromagnetic energy,
specifically from the dependence on the relative orientation of the
optical axes of the plates. The resultant torque tends to align
the optical axes along the configuration that minimizes the vacuum energy.

Recently, the Casimir effect has received considerable attention for
its possible technological applications, besides the fact that
experimental studies have attained the necessary accuracy to test in
detail the theoretical predictions
\cite{1,2,2a,2b,2c,2d,2e,3,3b,3c,3d,4,5,5b}. Therefore, 
theories about the Casimir effect that account realistically for the
properties of
actual materials have become indispensable.  The study of vacuum
forces between real materials was pioneered by Lifshitz \cite{9},
who considered two semi-infinite homogeneous and isotropic
non-spatially dispersive dielectric
slabs, whose fluctuating currents were the sources of the fluctuating
electromagnetic field and whose correlations were related to the
dielectric response of the materials. 

In 1972, Parsegian and Weiss derived an expression in the non-retarded
limit for the interaction energy between two
semi-infinite dissipationless dielectric anisotropic materials
\cite{10} following 
a method of surface mode summation.  Barash
derived an expression which included retardation and dissipation
effects \cite{11} employing an auxiliary system
\cite{barashginzburg} first introduced for isotropic systems. The solutions of
Maxwell's equations for the field in inhomogeneous and absorbing
media were expanded in terms of the orthogonal solutions of Maxwell's
equations for a non-dissipative auxiliary system in which the
frequency dependence of the dielectric function is only parametric.

The use of an auxiliary system was further developed in physical terms
by Kupiszewska \cite{kupi} in a calculation of Casimir forces for
lossy and dispersive isotropic dielectrics in the case of one
dimensional propagation of electromagnetic waves.  The
problem of quantizing a dissipative system is attacked by accounting
both for the dynamics of the vacuum modes and of the atomic dipoles to
which they couple and which make up the material, together with a
thermal reservoir in which the atomic radiators dissipate the absorbed
energy. That formalism was extended by van Enk \cite{12}
to obtain the torque between anisotropic materials in the 1D
case. In his work, van Enk
calculated the torque starting from the  flux of the spin angular
momentum of 
the electromagnetic field.
  
Numerical calculations have also been performed for materials with a
small anisotropy using Barash's results and it has been shown that the
torque may be large enough to be experimentally measurable in several
novel experimental configurations \cite{13}.  By a similar technique,
the Casimir energy between anisotropic dielectric plates \cite{14},
and between a plate with anisotropic magnetic response and another
with anisotropic dielectric response \cite{14a} 
have been calculated and analytical approximate expressions for the
torque and force were obtained in the retarded limit. Kenneth and
Nussinov have also calculated the Casimir   
energy between parallel plates made up of arrays of wires aligned
along different directions \cite{kenneth} using a path integral
technique \cite{kardar1,kardar2}.

In the  works described above, specific models of the dielectric
properties of the plates were assumed from the onset in order to
derive expressions for the Casimir force and torque. However, recent
works \cite{lambrecht,lambrecht-b,15,16,17,18,colnal} have shown that
if the theory is set up in 
terms of the reflection coefficients of the media, or equivalently, in
terms of their exact surface impedance \cite{21,22}, it is possible to
decouple the calculation of the Casimir force from the calculation of
the dielectric response of the materials.

Lambrecht et al. have extended their scattering approach
\cite{lambrecht,lambrecht-b} 
to corrugated systems  
\cite{72}. Moreover, they have argued that the resulting
formula for the Casimir energy has a
wider range of applicability and may be used to study other  
anisotropic mirrors. Their formula has been used to evaluate
numerically  the
effects of corrugation on the Casimir
force  \cite{96} and torque \cite{rodrigues} and 
to calculate the Casimir force between anisotropic metamaterials
\cite{milonni2}.

Moch\'{a}n {\em et al.} \cite{15,16,17,18,colnal} have argued that in thermal
equilibrium, all of the properties of the radiation field within a
cavity are completely determined by the optical reflection amplitudes
of the walls.  Indeed, whenever a photon reaches the surface of a wall
of the cavity 
it may be coherently reflected with an amplitude described by the
optical coefficients of the wall. Otherwise, it would be transmitted
into the wall to be either absorbed, exciting the material degrees of
freedom of its constituents, or transmitted across the wall and into
the surrounding vacuum to be lost forever. The probability of these
processes is again determined by the optical coefficients of the wall,
and given by their squared modulus. In thermodynamic equilibrium,
detailed balance implies that whenever a photon is lost, an equivalent
photon is incoherently injected back to the cavity, so that both, the
coherently and incoherently reflected photons are determined by the
reflection amplitudes alone. Therefore, the radiation field within a
real cavity would be identical to the field within any cavity that has
walls with the same optical properties. This fact allowed the
construction of fictitious disipationless systems which can be simply
treated quantum mechanically to obtain the vacuum energy and force for
cavities with arbitrary walls. 
Thus, expressions for the Casimir force obtained from
the electromagnetic stress tensor can be applied to semi-infinite
or finite, homogeneous or layered, local or spatially dispersive,
transparent or opaque systems through a simple substitution of the
appropriate optical coefficients. This formalism has allowed the
calculation of the Casimir force between photonic structures
\cite{23}, non-local excitonic semiconductors \cite{25},
non-local-plasmon-supporting metals with sharp boundaries
\cite{15,16,17,18,colnal,26}, and between realistic spatially
dispersive metals with a 
smooth self-consistent electronic density profile \cite{27}. With a
few modifications, it has also been employed for the calculation of
other macroscopic forces, such as those due to electronic tunneling
across an insulating gap separating two conductors \cite{29}.

The relative simplicity of the formalism developed in
\cite{15,16,17,18,colnal} has 
allowed its
generalization to anisotropic systems \cite{me}. In Ref. \cite{me}
a new derivation of the Casimir torque within  1D 
cavities with walls made up of arbitrary materials characterized only
by their anisotropic optical coefficients was presented. By 1D cavity
we mean one in which the field is constrained to propagate only along
one direction, namely, the normal to the surface of the cavity walls.
In the
present paper we 
generalize this formalism to 3D cavities with anisotropic walls. We calculate
the Casimir force from the electromagnetic stress tensor. A simple
integration over the separation distance $L$ yields then the vacuum
energy.
The torque is then calculated by 
taking the derivative respect to the angle $\gamma$ between the
optical axes of the plates. As our formalism is based on the
calculation of the force, which is a directly observable quantity, it
avoids the cumbersome singularities that plague other
approaches. Furthermore, our results are written directly in terms of
the optical coefficients of the walls of the cavity about which we
make no assumption. Thus, they can be applied immediately to manifold
systems such as insulating or conducting anisotropic slabs either
dissipationless or dissipative, to semiinfinite or finite walls and to
homogeneous or structured materials. We test the validity of our
approach by reproducing some known results
\cite{13,14,15,16,17,18,colnal,kenneth,lambrecht,lambrecht-b,milonni2} 
and we show its versatility by applying it to some previously
unexplored systems. 

The structure of the paper is the following: In section \ref{approach} we
develop our formalism in order to arrive at expressions for the
Casimir energy between 
arbitrary anisotropic plates. The use of an effective cavity allows
our results to be applicable to arbitrary slabs at any
temperature $T$. In section \ref{ideal} we specialize our results to
semi-infinite local uniaxial media and we apply our results to
a calculation of the energy and torque for an
idealized uniaxial system consisting of anisotropic mirrors that are
perfectly conducting along one direction and perfectly insulating
along the perpendicular directions. These calculations are performed
for temperatures $T=0$ and $T\ne 
0$. We also consider systems with a finite frequency-dependent
conductivity along the optical axis. As a further application of our
formalism, in section 
\ref{film} we calculate the torque of a 
system consisting of two conducting plates with an
anisotropic effective mass tensor, where one of the plates is a film
with finite thickness $d$, and we find there is an optimum value of
$d$ that maximizes the torque. We also calculated the same system
considered by Munday et al. \cite{13} consisting of a semi-infinite
slab of BaTiO$_3$ and a thin film of calcite. Finally, in section
\ref{conclusion} we present our conclusions. 
     
\section{The effective cavity approach}\label{approach}

\begin{figure}
\centering{
  \includegraphics{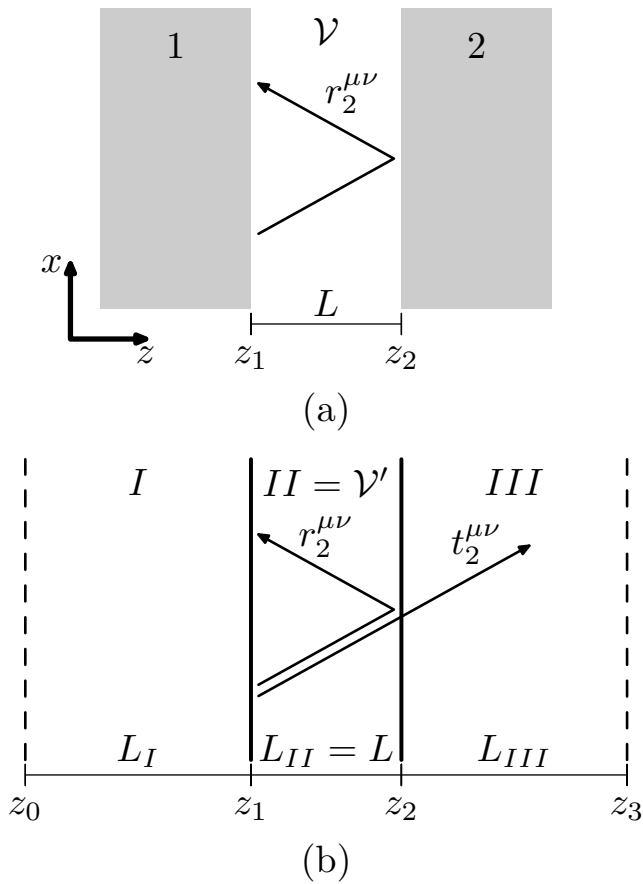}
}
\caption{\label{scattering} (a) Vacuum cavity $\mathcal V$ of width
  $L$ bounded by two arbitrary anisotropic material slabs (1 and 2)
  with surfaces at $z_1$ and $z_2$ and anisotropic reflection
  amplitudes $r_a^{\mu\nu}$ ($a=1,2$, $\mu,\nu=s,p$).
(b)Fictitious system made up three empty
regions $I$, $II$, and $III$, bounded by perfect mirrors at
$z_0$ and $z_3$ and with infinitely thin
sheets at $z_1$ and $z_2$ with identical 
reflection amplitudes 
$r_a^{\mu\nu}$ to those of the real system and corresponding
transmission amplitudes 
$t_a^{\mu\nu}$.}
\end{figure}

Consider the setup shown in Fig. \ref{scattering}(a). The slabs
represents arbitrary media.
According to \cite{15,16,17,18,colnal,30,me}, in thermodynamic equilibrium
the properties of the radiation field
within the cavity $\mathcal V$ are completely determined by the
geometry of the cavity, characterized by $L$, and by the $2\times 2$
reflection 
amplitude matrices $r_a^{\mu\nu}$ of each slab ($a=1,2$) coupling
$\nu$-polarized incident light to $\mu$-polarized reflected light
($\mu,\nu=s,p$). Any relevant property of the material is completely 
accounted for through its optical coefficients. Thus, the
electromagnetic radiation 
within the real cavity $\mathcal V$ must be identical to that
within a fictitious cavity $\mathcal V'=II$ bounded by infinitely thin
sheets at $z_1$ and $z_2$, provided their reflection amplitudes
$r_a^{\mu\nu}$ are chosen to match those of the walls of the real cavity
$\mathcal V$. The transmission amplitudes $t_a^{\mu\nu}$ of the
infinitely thin sheets are conveniently chosen in order to guarantee
energy conservation with no absorption whatsoever of electromagnetic
energy. Thus,
there is no absorption in the fictitious system, there is no
excitation of material degrees of freedom and the normal modes of its
electromagnetic field form a complete orthogonal basis of the
corresponding Hilbert space. As a consequence,
one is allowed to use well developed
quantum-mechanical procedures for the calculation of the field
properties without the requirement of a microscopic model of the material.

The field modes may be quantized and counted by choosing suitable
boundary conditions. For example, we can
can add perfect mirrors far away from the walls of the real
cavity (Fig. \ref{scattering}(b)),  at $z_0$ and $z_3$.  These
quantizing mirrors produce a 
field that mimics the incoherent radiation back into the cavity that
is responsible for maintaining a detailed balance and thus the
thermodynamic equilibrium.  

Consider now a single wave of frequency $\omega$ with wave-vector projection
$\vec Q$ along the interface. Without loss of generality, we choose
$x-z$ as the plane of incidence, so the electric and magnetic fields
are given by
\begin{equation}\label{E3D}
\vec{E}(\vec{r},t)=\mathcal{E}_{0} e^{i(Q x -\omega t)}
\left[\phi^s(z) \mathbf{\hat{y}} - \frac{1}{i q}( i Q \mathbf{\hat{z}}
  - \mathbf{\hat{x}}\partial_{z}) \phi^p(z)\right]%\nonumber\\ 
\end{equation}
and 
\begin{equation}\label{B3D}
\vec{B}(\vec{r},t)=\mathcal{E}_{0} e^{i(Q x -\omega t)}
\left[\phi^p(z) \mathbf{\hat{y}} + \frac{1}{i q}( i Q \mathbf{\hat{z}}
  - \mathbf{\hat{x}}\partial_{z}) \phi^s(z)\right], 
\end{equation}
where $q=\omega/c$ is the free-space wavenumber, $\phi^p(z)$ and
$\phi^s(z)$ are the $p$ and $s$ polarized components of a spinorial
normalized {\em 
  wave-function} 
\begin{equation}\label{phi}
\phi^\mu(z)= C^{\mu r}_\Lambda e^{ikz}+ C^{\mu l}_\Lambda
e^{-ikz}, 
\end{equation}
where $ C^{\mu\zeta}_\Lambda$ are constant coefficients  within each
region $\Lambda=I,II,III$ corresponding
to $\mu$-polarized light moving towards the right ($\zeta=r$) and left
($\zeta=l$), and
$k=\sqrt(\omega^2/c^2-Q^2)$ 
is the wave-vector component perpendicular to the interface. 

We integrate the energy density,  
\begin{equation}\label{u}
u=(|E|^2+|B|^2)/16\pi
\end{equation}
to obtain the total electromagnetic energy 
\begin{equation}\label{U}
\mathcal U=\frac{A|\mathcal E_0|^2}{8\pi} (L_I \, ||C_I||^2  +  L_{III} \,
||C_{III}||^2), 
\end{equation}
in the limit $L_I,L_{III}\to\infty$,
where $A$ is the area of the plates and $||C_\Lambda||^2 \equiv
\sum_{\mu\zeta} |C^{\mu\zeta}_\Lambda|^2$. In the same limit the
  normalization condition imposed on the wave-function simplifies to 
\begin{equation}\label{norm}
1= (L_I \, ||C_I||^2  +  L_{III} \, ||C_{III}||^2).
\end{equation}
Notice that most of the energy lies in the large fictitious regions I
and III, so we may identify  
\begin{equation}\label{U1}
\mathcal U=\frac{A|\mathcal E_0|^2}{8\pi}
\end{equation}
 and solve for the amplitude $|\mathcal E_0|^2=8\pi\mathcal{U}/A$.
To obtain the force on the slab 2, we calculate the stress tensor
$T_{ij}=(1/8\pi)\mathrm{Re}[E_i E_j^*+B_i B_j^*-(|E|^2+|B|^2)\delta_{ij}/2]$
at an arbitrary position $z$ within the cavity,
\begin{equation}\label{stensor}
-T_{zz}(z)=\frac{\mathcal U }{2Aq^2}
\left(k^{2}(\left|\phi^s\right|^{2}+
\left|\phi^p\right|^{2})+ \left|\partial_{z}\phi^s \right|^{2}+
\left|\partial_{z}\phi^p\right|^{2}\right).      
\end{equation}  

By applying boundary conditions at $z_0$ and $z_3$, we obtain for a
given value of $\vec Q$ a
discrete set of mode frequencies $\omega_n$ and corresponding
perpendicular components $k_n$ of the wave-vector. Each of these modes
contributes to the stress tensor
a quantity similar to that in Eq. (\ref{stensor}), so that
\begin{equation}\label{modes}
 -T_{zz}(z)=\frac{\hbar c}{2A}\sum_n
 \frac{f_n}{q_n} \bigl[k_n^{2} (| \phi^s_n |^{2} +
 |\phi^p_n  |^{2} )
+ (| \partial_{z} \phi^s_n
 |^{2}+  |\partial_{z} \phi^p_n |^{2} )\bigr]_z,\quad (\mathrm{fixed\ } \vec Q),
\end{equation}
where we have substituted the energy $\mathcal U_n$ in terms of the
equilibrium {\em occupation number}
$f_n=f(\omega_n)=\coth(\beta\hbar\omega_n/2)/2$ of a photon state with
quantized energy $\hbar\omega_n$ at temperature $k_B T=1/\beta$, with $k_B$
the Boltzmann's constant.  The sum over states may be rewritten in
terms of the tensorial Green's function $\mathbf{G}(z,z')$ with components
%\begin{equation}\label{G}
%\mathbf{\hat{G}}=\sum_{n}\frac{|n\left\rangle  \right\langle n
%|}{\tilde{k}^{2}-k_{n}^{2}}, 
%\end{equation}
%with matrix elements
\begin{equation}\label{Gmunu}
G^{\mu\nu}_{\tilde k^2}(z,z')=\sum_{n}\frac{\phi^\mu_n(z)
  \phi_{n}^{\nu*}(z')}{\tilde{k}^{2}-k_{n}^{2}},\; \mbox{ $\mu$, $\nu$
  = s, p,}
\end{equation}
for the 1D Helmholtz equation
\begin{equation}\label{eqG}
\left(\partial_{z}^{2}+ \tilde{k}^{2}\right)G^{\mu\nu}_{\tilde
  k^2}(z,z')=\delta(z-z')\delta_{\mu\nu}.
\end{equation}
Here, $\tilde{k}=k+i\eta$, ($\eta>0$), with the understanding that the
limit $\eta\to 0^+$ is to be taken at the end of the calculation.
Using the identity $\mbox{ Im}(\tilde{k}^{2}-k_{n}^{2})^{-1}=-\pi\delta(k^{2}-k_{n}^{2}) $, we can replace the sum
(\ref{modes}) by the integral
\begin{equation}\label{stotal}
-T_{zz}(z)=\frac{\hbar c}{A}\int dk^2\,
\frac{f}{q}\rho_{\tilde k^2},\quad (\mathrm{fixed\ } \vec Q),
\end{equation}
where $f=f(\omega)$ and
\begin{equation}\label{rho}
\rho_{\tilde k^2}(z)=-\frac{1}{2\pi} \mbox{ Im  }
\left(\tilde k^{2}+\partial_{z}\partial_{z'}\right)\left.\mathrm{Tr}\,
\mathbf{G}(z,z')\right|_{z=z'} 
\end{equation}
plays the role of a  local density of states at $z$ (number of
states per unit length and per unit $k^2$).

The solution of (\ref{eqG}), subject to the appropriate boundary
conditions, may be written in terms of the solutions
$\mathbf{u}(z)$ and $\mathbf{v}(z)$ of the 1D
Helmholtz equation that satisfy the boundary
conditions on the right and left side of the system, respectively,
\begin{eqnarray}
\mathbf{G}(z,z') &=& \mathbf{u}(z) \bigl[ \mathbf{u'}(z')
-
\mathbf{v}'(z') \mathbf{v}^{-1}(z') \mathbf{u}(z') \bigr]^{-1}
\theta(z-z')  \nonumber\\
&& -\mathbf{v}(z) \bigl[ \mathbf{v}'(z')
 - \mathbf{u}'(z')
\mathbf{u}^{-1}(z') \mathbf{v}(z') \bigr]^{-1} \theta(z'-z),
\label{Gs}
\end{eqnarray}
where  $\theta$ denotes the Heaviside unit step function. Here, 
$\mathbf{u}(z)$ and $\mathbf{v}(z)$ are $2\times2$ matrices with
matrix elements $u^\mu_\lambda(z)$ and $ v^\mu_\lambda(z)$,
$\lambda=1,2$ denotes the two independent spinorial solutions of
Helmholtz equation, $\mu=s,p$ denotes their correspondent $s$ and $p$
components and $\mathbf u'(z)$ and $\mathbf v'(z)$ denote the
derivatives of $\mathbf u(z)$ and $\mathbf v(z)$ with respect to their
argument. We remark that we first introduced a similar expression for
the spinorial Green's function in Ref. \cite{me}, where it was used to obtain
the angular momentum flux within a 1D cavity.

The solutions $\mathbf{u}(z)$ and $\mathbf{v}(z)$  may be written
within the cavity in
terms of the reflection coefficients of the plates  
\begin{equation}\label{rdef}
\mathbf r_a=\left( \begin{array}{cc}
r_a^{pp}&r_a^{ps}\\
r_a^{sp}&r_a^{ss}\end{array}\right), \quad (a=1,2).
\end{equation}
These are defined through
\begin{equation}\label{phitilde}
  \xi^\mu_{r2} = \sum_\nu r_2^{\mu\nu} \xi^\nu_{i2},
\end{equation}
where we define the unnormalized spinors $\xi^\mu_{i2}$ and
$\xi^\mu_{r2}$ ($\mu=s,p$) as the amplitudes of the incident and reflected
fields at the surface of plate 2 through 
\begin{equation}\label{ey}
E_y(z)=\xi^s_{i2}
e^{i\tilde k(z-L)} + \xi^s_{r2} e^{-i\tilde k(z-L)}
\end{equation}
 and
\begin{equation}\label{by}
B_y(z)=\xi^p_{i2} 
e^{i\tilde k(z-L)} + \xi^p_{r2} e^{-i\tilde k(z-L)}.
\end{equation}
The other components of the electromagnetic field may be obtained from
Eqs. (\ref{ey}), (\ref{by}) through Maxwell's curl equations.
The matrix elements
$r_1^{\mu\nu}$  of plate 1 are similarly defined.
Thus, we may write,
\begin{equation}\label{u3D}
\mathbf{u}(z)=\mathbf I
e^{i \tilde{k}(z-L)} +
\mathbf r_2 e^{-i \tilde{k}(z-L)}
\end{equation}
and
\begin{equation}\label{v3D}
\mathbf{v}(z)=\mathbf I e^{-i \tilde{k} z} +
\mathbf r_1 e^{i \tilde{k} z},
\end{equation}
where $\mathbf I$ is the  $2\times2$ unit matrix. 

Substitution of (\ref{u3D}) and (\ref{v3D}) in (\ref{Gs}), (\ref{rho})
and (\ref{stotal}) yields
\begin{equation}\label{flineal}
-T_{zz} = \frac{\hbar c}{\pi A} \mathrm{Re} \int dk^2 \, \frac{f\tilde
  k}{q\Delta } (1-e^{4i\tilde k L}r_{1}  r_2), \quad (\mathrm{fixed\ }\vec Q)
\end{equation}
where $r_a \equiv \mathrm{det}\,\mathbf{r}_a$ and 
\begin{equation}\label{Delta}
\Delta=\mathrm{det}\,(\mathbf{I}-e^{2i\tilde kL} \mathbf{r}_1
\mathbf{r}_2). 
\end{equation}
Note that $\Delta=0$ yields the dispersion relation of the  lossy modes of
the real cavity.
%and
%\begin{equation}\label{F}
%F(\mathbf{r}_{1},\mathbf{r}_{2})=
%\left(r_1^{pp}r_2^{pp}+r_1^{sp}r_2^{ps}+r_1^{ps}r_2^{sp}+r_1^{ss}r_2^{ss}
%\right). 
%\end{equation}
%Here, $\mbox{ det }$ is the determinant function,
%\begin{equation}
%$\mbox{ det }\mathbf{r}_{a}=r_a^{ss}r_a^{pp} -r_a^{sp}r_a^{ps}$.
%\end{equation}

Summing equation (\ref{flineal}) over $\vec Q$ we finally obtain
\begin{equation}\label{flsum}
-T_{zz}=\frac{\hbar c}{2\pi^3}\int d^2 Q \mbox{ Re }\int dk\, \frac{f
  k^2}{q\Delta }  (1-e^{4i\tilde k L}r_1 r_2),
\end{equation}
where we assumed Born-von Karman periodic boundary conditions along
the surface of area $A\to \infty$ to replace $\sum_Q\ldots\to A/(2\pi)^2\int
d^2 Q\ldots$
 
The flux of linear momentum $-T_{zz}$ in the fictitious cavity is the
same as in the real cavity between slabs 1 and 2. To obtain the force
on slab 2, we have to 
subtract the flux in the real system between the slab and
infinity. This can can be obtained following the same derivation given
above, 
but replacing the slab 1 by the complete system made up of slabs 1,
the cavity $\mathcal V$ and slab 2, and replacing slab 2 by empty
space. The result is identical to equation (\ref{flsum}), but
substituting $\mathbf{r}_2\to 0$. Thus, the total force per unit area
of slab 2 is
\begin{eqnarray}\label{force}
\frac{F_{z}}{A} &=& \frac{\hbar c}{2\pi^3} \mbox{ Re} \int d^2 Q \int
dk\,  f k^2 \frac{ e^{2i\tilde k L}}{q\Delta }
\nonumber\\
&&
\times\left(\mathrm{Tr}(\mathbf{r}_{1}\mathbf{r}_{2}) - 2e^{2i\tilde k L}
r_1 r_2\right)
\nonumber
\\
&=& -\frac{\hbar c}{4\pi^3} \mbox{Im}\int d^2 Q \int
dk\,  f \frac{ k}{q}
\frac{d}{d L} \log \Delta
\end{eqnarray}

The derivation above was performed for waves that propagate in vacuum,
that is, within the light cone $Q\le \omega/c$ and for real $k$. For
evanescent waves with 
$Q>\omega/c$ and imaginary $k$ the approach above has to be slightly
modified, as it turns to be impossible to choose fictitious
transmission amplitudes $t_a^{\mu\nu}$ that guarantee energy
conservation at the boundaries $z_1$ and $z_2$ of the fictitious
cavity of Fig. \ref{scattering}. Nevertheless, the fictitious system
may be easily altered to accommodate for evanescent waves \cite{30},
and it turns out that the expression (\ref{force}) and (\ref{energy})
remain valid even outside of the light cone \cite{30}. Thus, the
integration region of Eq. (\ref{force}) may include real and imaginary
values of $k$, as long as the wavenumber
$q\equiv\omega/c=\sqrt{Q^2+k^2}$ is real, i.e., $k$ should go along
the imaginary axis from $iQ$ to 0 and then along the real axis towards
infinity. 

The formalism developed above is a
generalization to anisotropic systems of the formalism developed in
Refs. \cite{15,16,17,18,colnal} for the isotropic case. The resulting
force (Eq. (\ref{force})) agrees with the usual Lifshitz's result
expressed in terms of the reflection amplitudes in the isotropic case
($r_a^{sp}=r_a^{ps}=0$).

%\section{The potential  energy and the torque}

The potential energy $U$ of the system may be now obtained by integrating
the force (\ref{force}) with respect to the plate separation from
$\infty$ towards the actual separation $L$, yielding
\begin{equation}\label{energy}
\frac{U}{A} = \frac{\hbar c}{4\pi^3} \mbox{ Im }  \int d^2 Q\int dk\,
f \frac{k}{q} \log{\Delta}.
\end{equation}
Notice that when
$L\to\infty$, $\Delta\to1$ as $e^{2i\tilde k L}\to 0$ for any positive of $\eta$.
We make a change of variable from $k$ to $q$ and
perform the usual rotation in the complex plane from the positive real
axis towards the 
imaginary axis to rewrite Eq. (\ref{energy}) as
\begin{equation}\label{energy1}
\frac{U}{A} = \frac{\hbar}{8\pi^3} \int_0^\infty du \int d^2 Q\,
\log{\Delta} 
\end{equation}
at $T=0$, where we introduce an imaginary frequency $\omega=iu$,
and as
\begin{equation}\label{energyqrotated}
\frac{U}{A}=\frac{k_B T}{4\pi^2} \mbox{ Re }\left.\sum_{\ell\ge 0}\right.' \int
d^2 Q  \log{\Delta_\ell}
\end{equation}
for $T\ne0$, where we have accounted for the poles of $f(iu)$ by
performing residue-like integrations at the Matsubara frequencies
$u_\ell=2\pi \ell k_B T/\hbar$ and we define
$\Delta_\ell=\Delta(\omega=i u_\ell)$. The
prime in the summation means that the $\ell=0$ term should be divided
by 2.
Eq. (\ref{energy1}) coincides with that
derived in Refs. \cite{lambrecht,lambrecht-b,milonni2} for $T=0$.

Notice that for anisotropic plates, the matrices ${\bf r}_a$ depend on
their in-plane orientation. Therefore, the energy $U$
(Eq. (\ref{energy})) depends implicitly on the relative
orientation $\gamma$ between 
the optical axes of the plates. 
Thus, we expect a torque $M$ on the plates 
which we may calculate simply by taking the 
derivative $M=-\partial U/\partial \gamma$.
It is easily verified that starting from Eq. (\ref{energy1}) but
setting $\vec Q=0$ instead of performing the integral $(A/4\pi^2)\int
d^2Q$ yields Eq. (13) of Ref. \cite{me}, i.e., the torque for a cavity
in which the field is constrained to propagate along only one
dimension, namely, along the normal to the surfaces. In ref. \cite{me}
the torque was obtained directly from the flux of angular momentum
within the cavity. Here, we took a different approach, obtaining the
torque from the angular dependence of the energy. The reason is that
the angular momentum is well defined only for finite width beams and
the overlap among the mulltiple reflections of a finite beam is
incomplete for oblique incidence and ill defined for evanescent
waves. 

We remark that our results are written in terms of the optical
coefficient matrices $\mathbf r_a$ of the walls of the cavity, about
which we have made no assumptions. Thus, our results may be applied to
systems of arbitrary absortance, conductivity and width, and they may
be homogeneous or inhomogeneous. Up to this point the dependence of
the energy, force and torque on the 
orientation of the plates has been implicit, through the unstated
dependence of $\mathbf r_a$. In the next sections we will apply our
result to specific cases where the dependence on $\gamma$ can be
exhibited explicitly.

\section{Semi-infinite uniaxial slabs}\label{ideal}

We consider semi-infinite uniaxial non-magnetic crystals with their
optical axes 
parallel to their surface. In this case, the 
reflection matrices $\mathbf r_1$ and $\mathbf r_2$ may be obtained as
particular instances for surfaces 1 and 2 of the formula \cite{Sosnowski}
\begin{equation}\label{r}
\mathbf r= (\mathbf s_2^{-1} +\mathbf s_1^{-1}\cos\theta
)^{-1}(\mathbf s_1^{-1}\cos\theta-\mathbf s_2^{-1}),
\end{equation}
which we derive in the appendix following the notation of
Ref. \cite{azzam}. Here,  $\mathbf s_1$ and $\mathbf s_2$ are the
$2\times 2$-matrices  
\begin{equation}\label{s1s2r}
\begin{array}{cc}
\mathbf s_1=\left(\begin{array}{cc}\frac{\tan\psi}{J}&
  -\frac{J^2\cot\psi}{In_o}\\ 1&
1\end{array}\right),& 
\mathbf s_2=\left(\begin{array}{cc}\frac{n_o^2\tan\psi}{J^2}&
  -\cot\psi \\ 1/J& 
n_o/I\end{array}\right),
\end{array}
\end{equation}
and
\begin{equation}\label{ij}
\begin{array}{ccl}
    I^2&=&n_o^2n_e^2- \sin^2\theta (n_o^2 \sin^2\psi+n_e^2 \cos^2\psi),\\
    J^2&=&n_o^2 -\sin^2\theta,
  \end{array}
\end{equation}
where $n_o$ and $n_e$
are the ordinary and extraordinary refractive indices of the uniaxial
crystal, $\theta$ is the angle of incidence and $\psi$ is the angle
from the plane of incidence to the optical axis of the crystal;
its sign is chosen through the right hand rule around the normal of
the surface that points outwards from the anisotropic medium.
In Eqs. (\ref{s1s2r})  we have assumed that the cavity is empty and
thus we took its index of
refraction of the cavity as 1.
\begin{figure}
  \centering
  \includegraphics[width=.45\textwidth]{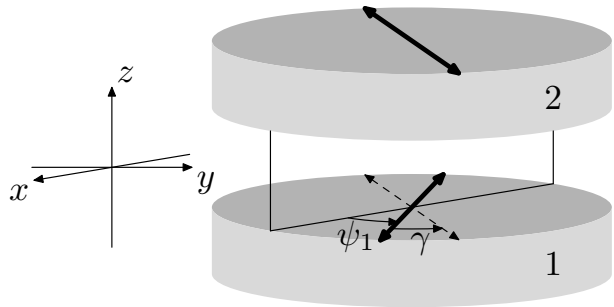}
  \caption{\label{config} Schematic diagram of the system, consisting
    of two parallel uniaxial non-magnetic plates whose optical axes
    lie on the surface. We indicate the optical axes of each plate
    with heavy double-headed arrows, and the projection of the axis of
    the second plate upon the first by a dashed double arrow. The
    $z$-axis is chosen to be orthogonal to the
    plates. We also indicate the plane of incidence. The angle
    $\psi_1$ between the plane of incidence and the optical axis of
    plate 1 as well as the angle $\gamma$ from the optical axis of
    plate 1 towards that of plate 2 are indicated. The angle $\psi_2$
    between the plane of incidence and the optical axis of plate 2 is
    $\psi_2=-\psi_1-\gamma$. The sign is due to the convention in
    Eq. (\ref{s1s2r}).
  }
\end{figure}

Consider the sketch depicted in Fig. \ref{config}, which displays
the  
angle $\psi_1$ of the optical axis of plate 1 with respect to the
plane of incidence and the angle $\gamma$ between the optical axis of
plate 2 and that of plate 1. According to our convention above, the
angle $\psi_2$ between the optical axis of plate 2 and the plane of
incidence is $\psi_2=-\psi_1-\gamma$. We identify
$n_e=\sqrt \epsilon_{\|}$, $n_o=\sqrt\epsilon_{\perp}$, where
$\epsilon_{\|}$ and $\epsilon_{\perp}$ are the 
dielectric response functions of the plates along and perpendicular to
their 
optical axes, respectively. The
substitution of equations (\ref{s1s2r}), (\ref{ij}) in (\ref{r}) and
(\ref{energyqrotated}) yields after a tedious algebra an expression
for the Casimir energy between anisotropic plates. We have checked
that the resulting expression for $\Delta$ coincide with that obtained
in 
Ref. \cite{14} for the same system.
%We have compared 
%our results with those obtained in
%Ref. \cite{14} for the same system, and we have
%verified that they coincide. 
Nevertheless, the calculation of
Ref. \cite{14} is directly
applicable only to local uniaxial semi-infinite homogeneous systems, as
their calculation is setup in terms of the bulk dielectric function of
the 
plates. On the other hand, our result is written in terms of the
reflection amplitudes $\mathbf r_a$ of the plates, and thus may be
applied to any system for which we can calculate these optical
coefficients, as shown explicitly in section \ref{film}.

Let us now consider an idealized case consisting of slabs which behave
as perfect conductors along the optical axes but which are perfect
insulators along the other principal directions. This could be realized through
an array of perfectly conducting aligned wires insulated from each
other. We further ignore the electric polarization across the wires,
which would be appropriate, for example, if they were very thin. Then, 
each plate may be described by a dielectric response 
$\epsilon_{\|}\to\infty$ and $\epsilon_{\perp}\to 1$. In this
case the reflection matrix (\ref{r}) becomes simply
\begin{equation}\label{rideal1}
\mathbf{r}(\theta,\psi)\!=\!\left[\!\begin{array}{cc}
\cos^{2}\theta \cos^{2}\psi&-\sin2\psi  \cos\theta/2\\
\sin2\psi \cos\theta/2&-\sin^{2}\psi\\
\end{array}\!\right]\!\left(1-\sin^{2}\theta \cos^{2}\psi \right)\!^{-1},
\end{equation}
which yields
\begin{equation}\label{Deltaideal}
  \Delta=1-\alpha^2 e^{2 i k L}
\end{equation}
when substituted into Eq. (\ref{Delta}),
where 
\begin{equation}\label{alpha2}
  \alpha^2 = \mathrm{Tr}(\mathbf r_1 \mathbf r_2)
  =\frac{\left( \cos\gamma - \sin^2 \theta \cos \psi_1 \cos\psi_2
    \right)^2} {(1 - \sin^2\theta \cos^2\psi_1) (1 - 
    \sin^2\theta \cos^2\psi_2)}
.
%  =\frac{\left( \cos\gamma - \sin^2 \theta \cos \varphi \cos(\varphi -
%    \gamma) \right)^2} {(1 - \sin^2\theta \cos^2\varphi) (1 -
%    \sin^2\theta \cos^2(\varphi-\gamma))},
\end{equation}
Eqs. (\ref{Deltaideal})
and (\ref{alpha2}) were previously obtained for this system in 
Ref. \cite{kenneth}, as becomes 
evident by writing the latter in 
terms of $\varphi\equiv -\psi_2 = \psi_1+\gamma$. 

\begin{figure}
  \includegraphics[width=.45\textwidth]{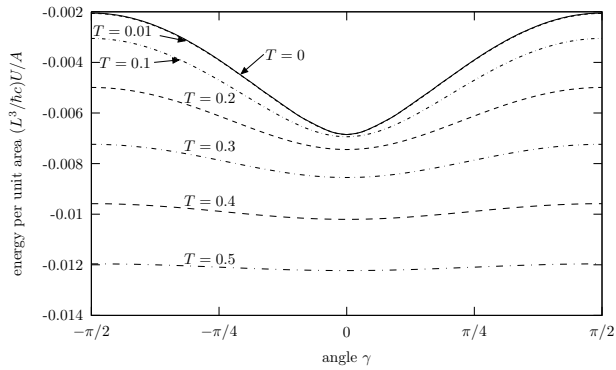}
  \caption{\label{energyvsa}Casimir energy per unit area  $U/A$
    %in units of $\hbar c/L^3$ 
    between two ideal uniaxial
    plates as a function of the angle $\gamma$ between their optical
    axes for different temperatures $T=0,0.01, 0.1, 0.2, 0.3, 0.4, 0.5$
    in units of $\hbar c/L k_B$. } 
\end{figure}

Substitution of Eq. (\ref{Deltaideal}) into
Eq. (\ref{energy1}) yields the $T=0$ energy of the system. 
Fig. \ref{energyvsa} shows our numerical results for the energy of
the idealized system discussed above. We have 
verified that they agree with the results presented in
Ref. \cite{kenneth} where a path integral  
approach \cite{kardar1,kardar2} was used.  
Nevertheless, a similar substitution into
Eq. (\ref{energyqrotated}) permits us to calculate also the $T\ne 0$
energy of the system, 
shown in the same figure. Notice that for this idealized case
there is a natural temperature scale $\hbar c/k_B L$ and a natural
energy scale $A \hbar c/L^3$, as they are respectively intensive and extensive
quantities with respect to the area $A$ of the interfaces.  We use
these scales to normalize the units in the
figure. As $T$ increases, the Casimir 
energy becomes more negative, so that the plates are more strongly
bound. However, the slope of the energy as a function of the angle, 
and thus, the torque, diminishes with increasing $T$. This
result may appear somewhat surprising, as it is well known that increasing the
temperature yields an increase of the magnitude of the Casimir
force.  For  high enough 
temperatures the energy becomes independent of the angle and
approaches its value 
at $\gamma=0$, which is exactly half of the energy corresponding to
two perfect isotropic mirrors. This can be easily verified
by substituting $\alpha^2=1$ into Eq. (\ref{Deltaideal}) and the
resulting $\Delta$ into (\ref{energy1}) and \eqref{energyqrotated}. 

The torque for the ideal case may now be simply calculated by
analytically deriving under the integral sign
the energy (\ref{energy1}) or (\ref{energyqrotated}) with 
respect to the relative angle $\gamma$ and performing the resulting
integrals numerically.
\begin{figure}
 \begin{center}
 \includegraphics[width=.45\textwidth]{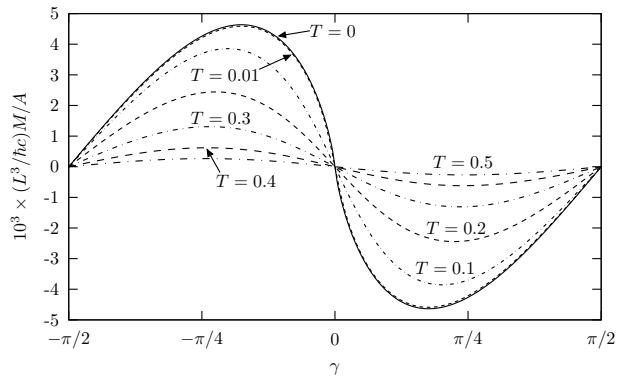}
 \end{center}
  \caption{\label{tvsaideal}Casimir torque  per unit area $M/A$
    %in units of $\hbar c/L^3$ 
    between two ideal uniaxial
    plates as a function of the angle $\gamma$ between their optical
    axes. The torque is calculated for the same temperatures as in 
    Fig. \ref{energyvsa}, expressed in units of $\hbar c/L k_B$. } 
\end{figure}
Fig. \ref{tvsaideal} shows that the torque $M$ is a periodic function
of $\gamma$ 
with period $\pi$.  It is null when the optical axes are aligned,
$\gamma=0,\pi$, corresponding to a stable equilibrium orientation. It is
also null when they are orthogonal to each other, $\gamma=\pm\pi/2$,
corresponding to an unstable equilibrium.  For $T=0$ the slope
of 
$M(\gamma)$ seems singular at the stable equilibrium point. The
existence of this singularity can actually be confirmed by deriving
analytically the torque with respect to
the angle $\gamma$ under the integral sign and examining the integrand
after having taken the corresponding limit
$\gamma\to0$ and
analytically performed the angular integral over $\varphi$. It turns
out that the 
integrand has an infinite discontinuity at $u=0$. 
This singularity had also been found previously in a 1D calculation
\cite{me}, in which light is normally incident at the walls of the
cavity. Furthermore, qualitatively the same behaviour of the torque as
a function of the angle $\gamma$ found in this 1D calculation, is
obtained in the 3D calculation for $T=0$. The torque is not simply
proportional to $\sin 2\gamma$, so its extreme values are
not at $\gamma=\pm\pi/4$. However, as the temperature increases the
torque becomes more sinusoidal-like.
Note that the maximum torque per unit area is of the
order of $M/A\sim 5\times 10^{-3}\times \hbar c/L^3$. Thus, for a
separation $L=100$ 
nm, the maximum torque per area unit is about $10^{-7} $N$/$m. It
is interesting to compare this result with that of  Rodrigues et
al. \cite{rodrigues} for the case of two slightly misaligned
corrugated metals. It has been argued \cite{rodrigues} that that system
would yield the largest torques. Nevertheless, we obtained a torque of
the same order of magnitude as their's. Furthermore,
in our case the torque remains large over a wide angular range,
independently of the area of the system, while in
their case the torque is significant only within a very small angular
range which decreases with the size of the system. 

\begin{figure}
 \begin{center}
 \includegraphics[width=.45\textwidth]{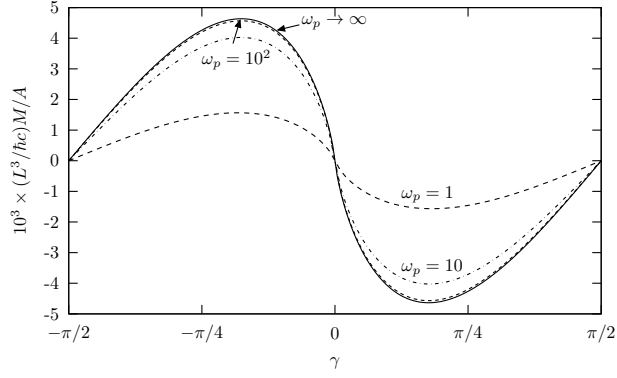}
 \end{center}
  \caption{\label{tvsanoideal}Casimir torque at $T=0$ between two uniaxial
    plates as a function of the angle $\gamma$ between their optical axes.
    The plates are conducting along their optical axis with a
    response $\epsilon_\|$ described by a Drude dielectric
    function with a 
    relaxation time  $\tau$, 
   and a plasma frequency $\omega_p$, and we assume the response
   perpendicular to the optical axis is $\epsilon_\perp=1$. We chose a
   fixed value 
   $\tau=1000/\omega_0$,  and we show results for various values of
   $\omega_p=1$, $10$ and $10^2\times \omega_0$ where
   $\omega_0= \pi c/L$. We also include the case $\omega_p=\infty$
   corresponding to the ideal case studied above. 
 } 
\end{figure}

Our theory may be applied to systems which are more realistic than the
idealized case above. For example, 
in Fig. \ref{tvsanoideal} we show the zero temperature torque between
two uniaxial conductors whose response along the optical axis is characterized 
by the Drude  dielectric function
\begin{equation}\label{Drude}
  \epsilon_{\|}(\omega)=1 - \frac{
  \omega_{p}^{2}} {\omega^2+i\omega/\tau}.
\end{equation}
For simplicity, we assume they do not respond along the perpendicular direction,
$\epsilon_\perp=1$, as could correspond to parallel plates made up of
an array of very 
thin aligned wires, each of which is a conductor with a finite
Drude-like conductivity.
We include results for different values of the electronic density and
therefore of the plasma frequency $\omega_p$, as well as the
ideal limit $\omega_p\to\infty$. In the figure we used
$\omega_0=\pi c/L$ as a natural frequency scale and we chose the
electronic lifetime $\tau=10^3/\omega_0$ which would correspond  to a typical
values within a metal for $L\sim 10^2$nm, although the results are
very insensitive to $\tau$. 
In this case, $\omega_0$
would be of the order of a typical metallic plasma frequency. 

When $\omega_p\gg \omega_0$ the plates
behave as perfect mirrors and, as verified by Fig.
\ref{tvsanoideal}, we recover the idealized case studied above,
including the approach to the singularity in the slope at
$\gamma=0$. As expected, the torque increases with
$\omega_p$ and becomes null at $\omega_p=0$.

\section{Films}\label{film}

Our formalism, summarized by Eqs. (\ref{energy1}) and
(\ref{energyqrotated}), is written entirely in terms of the reflection
amplitudes $\mathbf r_1$ and $\mathbf r_2$ of the anisotropic plates
through Eq. (\ref{Delta}). This has the enormous advantage over
previous formalisms \cite{barashginzburg,14,14a} in that
different systems may be explored simply by substituting their
corresponding reflection amplitudes. For arbitrary uniaxial systems we
could simply substitute the ordinary and extraordinary indices of
refraction in the formulae derived in the previous section. 
In this section we further illustrate the
versatility of our theory by calculating 
the torque between
two anisotropic dielectric plates one of which is
a film with finite thickness $d$.  For the reflection amplitudes
$\mathbf r_1$ of the semi-infinite plate, which we take to be plate 1,
we can simply use Eq. (\ref{r}), but we have to calculate
the  
reflection coefficients $\mathbf r_2$ of the film. 
 
Consider the reflection of an arbitrary polarized plane-wave with
frequency 
$\omega$ from an
uniaxial anisotropic free-standing homogeneous dielectric film of finite
thickness $d$, with its optical axis parallel to its surface.   
We assume that the system is described by the same geometry as in the
previous section, illustrated by Fig. \ref{config}, and that the
film is non-magnetic. 

For each frequency $\omega$ and parallel projection  $\vec Q$ of the
wave-vector, two types of waves coexist inside the film, the $ordinary$ and
$extraordinary$ waves, each of which has a wave vector component
$\pm k_\mu$ normal to the surface, where $\mu=o$ for ordinary waves,
$\mu=e$ for extraordinary waves and we choose the upper sign for waves
travelling or decaying towards $z$, i.e., we choose $k_\mu$ as the
solution of the dispersion relation that is consistent with 
$\mbox{Im}(k_\mu)>0$ for evanescent waves and for propagating waves in
the presence of finite or infinitesimal absorption.
This gives a total of four waves in the film. Thus, we describe the
electromagnetic field within the film as \cite{azzam}
\begin{equation}\label{bmpsi}
  \bm F(z) = \sum_{\sigma,\mu} \psi_\sigma^\mu \bm V_\sigma^\mu
  e^{i\sigma k_\mu z},
\end{equation}
where $\bm F$ is a 4-component column vector which contains the
electric and magnetic field projections parallel to the surface, $\bm
F = (E_x, B_y, E_y, -B_x)^T$, $\psi_\sigma^\mu$ is the amplitude of
the normal mode propagating in the $\sigma z$ 
direction ($\sigma = +,-$) with polarization $\mu=o,e$, while
$\bm V_\sigma^\mu$ is the eigenvector describing the electromagnetic
field $\bm F$ of a single mode with propagation direction $\sigma$ and
polarization $\mu$, and  $\sigma k_\mu=\pm k_\mu$ is the  component of
the corresponding wave-vector along $z$ (see the appendix for details). 

The field reflected by the film has two contributions. One of them is
the field reflected by its front surface. The other is the field
transmitted back into the cavity from the film, after having entered
the film from the cavity and being multiply reflected. Thus,
we may write 
\begin{equation}\label{frontface1}
\mathbf r_2 \bm\xi_{i2}=\mathbf r_{02}\bm \xi_{i2}
+\mathbf t_{20} \bm \psi_- 
\end{equation}
where $\bm
\psi_\sigma=(\psi_\sigma^o,\psi_\sigma^e)^T$ is a 2-component column
vector containing 
the total $o$ and $e$ amplitudes of the waves travelling in the
$\sigma z$-direction within the film, $\bm\xi_{i2}
=(\xi_{i2}^p,\xi_{i2}^s)^T$ is the vector containing the
$p$ and $s$ contributions of the incident field and is 
defined through Eqs. (\ref{ey})-(\ref{by}), 
the matrix $\mathbf r_2$ is the
sought reflection amplitude of the film, $\mathbf r_{02}$ is
the reflection amplitude of the front surface when light impinges from
the cavity and therefore it is
given by Eq. (\ref{r}),
and  $\mathbf t_{20}$ is the transmission amplitude from the
film into the 
cavity across the front surface. Similarly, the field that travels
towards $z$ within the film is the sum of the field transmitted from
the cavity into the film plus the field travelling towards $-z$ and
internally reflected back at the front surface, so
we may write
\begin{equation}\label{frontface2}
\bm \psi_+=\mathbf t_{02} \bm \xi_{i2}+\mathbf r_{20} \bm \psi_-,
\end{equation}
%Therefore,
%$\mathbf r_{02}$ is given 
%by Eq. (\ref{r}). 
where $\mathbf t_{02}$ is the transmission amplitude from the cavity
into the film through the front surface, and $\mathbf r_{20}$ is the 
internal reflection amplitude of the front surface when light impinges
from within the
film.
Similarly, at the rear surface, we may write
\begin{equation}\label{rearface1}
\bm\kappa^{-1} \bm \psi_-=\mathbf r_{20}\bm \kappa\bm \psi_+
\end{equation}
where  $\bm \kappa=\mbox{diag}(e^{ik_od},e^{ik_ed})$ is a 
$2\times 2$ diagonal matrix that accounts for the phase acquired by the
 ordinary and extraordinary waves as they travel from the
 front surface towards the rear surface of the film. Notice that the
 internal reflection matrix of the rear interface  
 coincides with the internal reflection matrix at the front interface as we
 consider a free standing film, and hence we
 denote both matrices by  
 $\mathbf r_{20}$. Elimination of $\bm
 \psi_+$ and $\bm \psi_-$ from 
Eqs. (\ref{frontface1})-(\ref{rearface1}) yields  
\begin{equation}\label{rfilm}
\mathbf r_{2}=\mathbf r_{02}+\mathbf t_{20}\bm \kappa \mathbf
r_{20} \bm \kappa (\mathbf I -(\mathbf r_{20}\bm \kappa)^2
)^{-1}\mathbf t_{02}.
\end{equation}
We remark that the same result would be obtained by generalizing
Airy's method \cite{airy},
summing the amplitudes of successive multiple
reflections and
refractions, but taking account of their $2\times2$-tensorial
character due to the mixing of polarizations at the
interfaces. Equivalent results may also be obtained through a
$4\times4$ transfer matrix formalism \cite{berreman,azzam}, though
care must be taken to avoid numerical instabilities.
The calculation of the required matrices $\mathbf t_{02}$, $\mathbf
t_{20}$, 
and $\mathbf 
r_{20}$,  as well as the calculation of the wave vector components
$k_\mu$ is shown in the Appendix.  

By substituting Eq. (\ref{rfilm}) into Eqs. (\ref{energy1})
or (\ref{energyqrotated}) we finally obtain the Casimir energy
corresponding to an anisotropic film in front of a semi-infinite,
anisotropic 
substrate. Then, the torque is obtained as before, simply by deriving
analytically the energy under the integral sign with respect to the angle
$\gamma$ between the optical axes. 

\begin{figure}
 \begin{center}
\includegraphics[width=.45\textwidth]{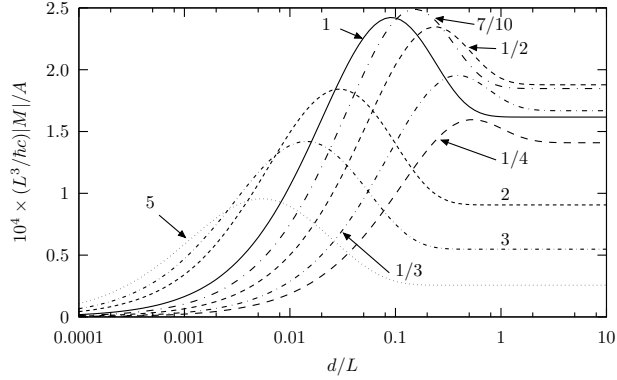}
 \end{center}
  \caption{\label{tvsdwpe_wpo2}Casimir torque at $T=0$ between a
    semi-infinite uniaxial, dissipation-less
    conducting slab and a film made up of the same  
    material as a function of thickness $d$ of the film. The slabs 
    are modelled as Drude metals with a plasma frequency
    $\omega_{p\parallel}$ along the optical axis twice as large than
    its value $\omega_{p\perp}$ along the perpendicular direction. The
    curves are labelled by the value of 
    $\omega_{p\perp}/\omega_0=1/4,1/3,1/2,7/10, 1,2,3,5$, where
    $\omega_0=\pi c/L$. 
    We fixed the relative angle between the optical 
    axes at $\gamma=\pi/4$.}
\end{figure}

In Fig. \ref{tvsdwpe_wpo2} we show the torque between a semi-infinite
anisotropic substrate and an anisotropic film  as a function of the
thickness $d$ of the film. The substrate and the film are made of the
same material which we assume is an uniaxial conductor with
different plasma frequencies along its principal directions,
i.e., with a tensorial anisotropic effective mass, and which we model
by a Drude-like response along and perpendicular to the optical
axis  
\begin{equation}\label{Drudeboth}
\epsilon_{\mu}(\omega)=1- \frac{\omega_{p\mu}^{2}}
{\omega^2+ i\omega/\tau},\quad \mu=\perp,\parallel.
\end{equation}
We include results for
different values of $\omega_{p\perp}$ for a fixed quotient 
$\omega_{p\parallel}/\omega_{p\perp}=2$ and for simplicity, we
disregarded the dissipation, assuming
$\tau=\infty $.  

 From Fig. \ref{tvsdwpe_wpo2} we see that for very thin films the
 torque 
 becomes smaller as the normalized film
 thickness $d$ becomes smaller and eventually the torque becomes null
 for 
 a  zero film thickness $d\to0$, as $\bm r_2\to0$ vanishes in that 
 limit, as can be verified from Eq.  (\ref{rfilm}). For large enough
 values 
 of $d$ the torque tends to its asymptotic value
 corresponding to semi-infinite plates. 
Curiously, the figure shows optimal  values of the film thickness for
which 
the  torque is maximized. For the chosen parameters, 
the maximum torque may be as large as twice its asymptotic value. The
optimal thickness shifts towards 
smaller values as the plasma frequency $\omega_{p\perp}$
is increased.  Furthermore, the
magnitude of the torque at the optimal thickness  
is maximum for values of $\omega_{p\perp}$ around $\omega_0$. 

In order to explain qualitatively the results
shown in Fig. \ref{tvsdwpe_wpo2}, we consider a 1D model in which 
light is normally incident at the walls of the cavity. In this case,
it is possible to decouple the ordinary and extraordinary rays inside the
media by choosing the polarization of the incident electrical field. 
Light polarized in the directions normal or parallel to the optical
axis will only excite the ordinary or extraordinary wave inside
the media and will be reflected without changing its polarization with
reflection amplitudes $r^\perp$ or $r^\parallel$ respectively. The
torque at $T=0$ for this 
one dimensional system can be calculated using the results of Ref.
\cite{me},  
\begin{equation}\label{taufin}
    M=-\frac{\hbar \sin 2\gamma}{2\pi}\int_0^\infty du
    \frac{\Delta r_1 \Delta r_2 \,e^{-2uL/c}}
  {
      \begin{array}{l}
	\Delta r_1 \Delta r_2\sin^2 \gamma e^{-2uL/c}
	+(1-r_1^\parallel r_2^\parallel e^{-2uL/c})
	(1-r_1^\perp r_2^\perp e^{-2uL/c})
      \end{array}
  },
\end{equation}
where $\Delta r_a=r_a^\parallel-r_a^\perp$.
The  reflection  coefficients are \cite{wolf}
\begin{equation}
r_1^\mu=\frac{1-\sqrt{\epsilon_\mu}}{1+\sqrt{\epsilon_\mu}}
\end{equation}
for plate $1$,  and \cite{wolf}
\begin{equation}\label{scalarr}
r_2^\mu=r_1^\mu \frac{1-
  e^{2ik_\mu d}} {1 -(r_1^\mu)^2e^{2ik_\mu d}}
\end{equation} 
for the film, where $\mu=\parallel,\perp$ and
$k_\mu=\sqrt{\epsilon_\mu}\omega/c$ is the wave number inside the
film corresponding to $\mu$-polarization. 

According to Eq. (\ref{taufin}), we might understand the dependence of
the torque $M$ on the thickness $d$ of the film if we first understand the
dependence of the anisotropy $\Delta r_2$ on $d$. Consider first a wave of
frequency $\omega>\omega_{p\perp},\omega_{p\|},$ normally incident on
the film 2. In this case, the film would be transparent for
both polarizations and the anisotropy
would be small. For intermediate frequencies, $\omega_{p\perp}<\omega
<\omega_{p\|}$, the film 
would be a good reflector for one polarization and transparent for the
perpendicular polarization, yielding a large anisotropy which
nevertheless would be quite insensitive to the thickness of the
film. However, for $\omega<\omega_{p\perp} < \omega_{p\|}$ 
the penetration depth $\ell_\perp$ for polarization perpendicular
to the optical axis would be larger than the penetration depth $\ell_\|$ for
polarization along the optical axis. For a sufficiently thick film,
$d>\ell_\perp > \ell_\|$ and the film 
would be a good reflector for both polarizations as no radiation would
get across it, so the anisotropy would be small. For a sufficiently
thin film $\ell_\perp>\ell_\|>d$, the radiation would reach the back
surface and leave 
the film, which would be a poor reflector for both
polarizations and thus the anisotropy would again be
small. In the limit of vanishing thickness $d\to 0$
all the incident radiation would be transmitted through the film and
none reflected. 
Nevertheless, for an intermediate thickness such that
$\ell_\perp > d > \ell_\|$ the film would be a good reflector for 
polarization along the optical axis but a poor reflector along the
perpendicular direction, yielding a large anisotropy.

\begin{figure}
 \begin{center}
\includegraphics[width=.45\textwidth]{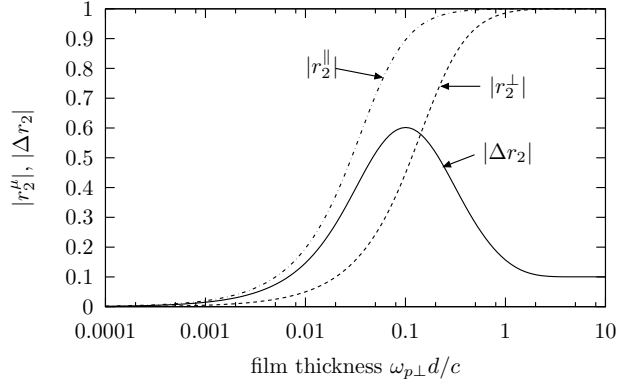}
 \end{center}
  \caption{\label{skineff} Magnitude of the
 normal-incidence reflection coefficients $|r_2^\mu|$ of an anisotropic, thin,
 dissipation-less Drude conducting
 film with plasma frequencies $\omega_{p\|}=2\omega_{p\perp}$, and
 magnitude of the reflection anisotropy $|\Delta 
 r_2|$ as a function of the film thickness $d$. The frequency is
 $\omega=0.1\omega_{p\perp}$. } 
\end{figure}

These qualitative features are displayed in Fig. \ref{skineff}, where
we show the magnitude of the reflection coefficients $|r_2^\mu|$ as well as
the magnitude of the anisotropy $|\Delta r_2|$ as a function of 
$d$ for the case $\omega=0.1\omega_{p\perp}$. In the limit of large
thickness $d\gg c/\omega_{p\perp}$, the anisotropy of the film $\Delta r_2$
tends to its asymptotic value, which is non zero due to the
relative  phase between the complex optical coefficients corresponding to
both polarizations.  The film thickness $d^m$ which maximizes the
anisotropy can be calculated from Eq. (\ref{scalarr}) in a
straightforward way, yielding 
\begin{equation}\label{kpoptimo}
d^m\approx
\frac{2|\omega|c}{\omega_{p\perp}\omega_{p\parallel}}, \quad
|\omega|<\omega_{p\perp},\omega_{p\|} 
\end{equation}
for small real or imaginary frequency.

\begin{figure}
 \begin{center}
\includegraphics[width=.45\textwidth]{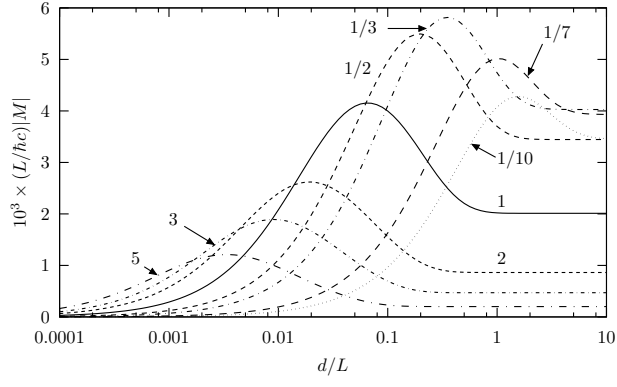}
 \end{center}
  \caption{\label{t1Dvsdwpe_wpo2}Casimir torque 1D (only light
    normally incident is accounted for) at $T=0$ of the
    same system considered in Fig. \ref{tvsdwpe_wpo2}, as a function
    of the  thickness of the film $d/L$. The torque is
    calculated for $\omega_{p\perp}=\omega_p=1/10,1/7, 
    1/3, 1/2,1,2,3,5$ (in units of $\omega_0=\pi c/L$).  The rest of
    the parameters are the same as in Fig. \ref{tvsdwpe_wpo2}.
  }
\end{figure}

In Fig.  \ref{t1Dvsdwpe_wpo2} we show the torque as a function of the
width $d$ for the same system as in Fig. \ref{tvsdwpe_wpo2} but
assuming that the field propagates only in 1D \cite{me}. The results display  the
same qualitative behavior as in the 3D case,
although  the plasma frequency
$\omega_{p\perp}$ for which the 
magnitude of the torque attains its maximum is shifted to a smaller value
$\omega_{p\perp}/\omega_0\approx 1/3$ compared to $7/10$ in
Fig. \ref{tvsdwpe_wpo2}. We may estimate the optimal width for each
value of $\omega_{p\perp}$ through the following considerations:
The exponential in  
Eq. (\ref{taufin}) suppresses the contribution of large imaginary
frequencies $u> c/2L$  to the Casimir effect. Thus, in the retarded regime 
$\omega_{p\perp} > \omega_0$ we may consider  mostly low imaginary frequencies
$u/\omega_{p\mu}<1$  to understand the effect. Furthermore, as  the 
relevant frequency scale is $c/L\sim \omega_0/\pi$, then 
the localization of the optimal film thickness in the retarded
regime could be estimated by substituting $|\omega|=u=\omega_0/\pi$
within  Eq. (\ref{kpoptimo}). 
Thus, we obtain  $d^m/L\approx 0.025$, $0.01$ and $0.004$ for the choices
$\omega_{p\perp}/\omega_0=2$ , $3$ and $5$
respectively. These estimates are in good agreement with the 1D
calculation, Fig. \ref{t1Dvsdwpe_wpo2}. In the 3D case one would have
to take into account modes that propagate along non-normal
directions. However, as the reflection amplitude increases towards
unity for any polarization as the angle of incidence moves away from
the normal direction, the anisotropy in the reflectance and
the corresponding contribution to the torque also diminish. Thus, the
most important contributions to the torque come from modes that
propagate close to the normal and our estimates for $d^m$ are also in
good qualitative agreement with our full 3D calculation,
Fig. \ref{tvsdwpe_wpo2}. 

The simple estimates above fail in the
non-retarded regime $\omega_{p\perp}  
< \omega_0$ for both the 1D and 3D calculation, as in this case a
wider range of frequencies, going beyond
$\omega_{p\perp},\omega_{p\|}$, contributes to the
torque. Furthermore, for small separations $L$ the Casimir effect 
in the 3D case is dominated by surface plasmons \cite{Intravia} which
are completely left out of the 1D calculation and of our estimates
above.

\begin{figure}
 \begin{center}
\includegraphics[width=.45\textwidth]{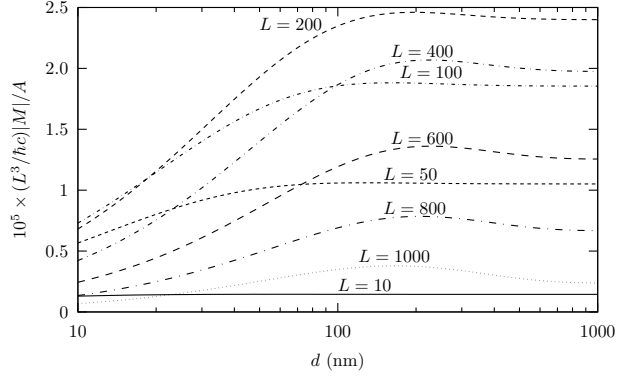}
 \end{center}
  \caption{\label{iantfvsdvsL}Casimir torque at $T=0$ between a
    semi-infinite substrate of Ti$\mbox{O}_3$ and a film of calcite,
    as a function of the  thickness of the film $d$. The torque is
    calculated for several separations $L=10, 50, 
    100, 200,400,600,800,1000$ nm.  The angle between the optical axes 
    is $\gamma=\pi/4$.
  }
\end{figure}

To conclude this section, in Fig. \ref{iantfvsdvsL}  
we show the Casimir torque between a crystal of BaTiO$_3$ and a
crystal of calcite, as in Ref. \cite{13}. However, instead of two
semi-infinite crystals, we consider a thin calcite film over a
semi-infinite BaTiO$_3$ substrate with vacuum in between and we vary
the thickness $d$ of the film. We use the same model  as in
Ref. \cite{13}  to describe the 
dielectric response of the plates along each of their  principal
directions, namely, two undamped oscillators \cite{Bergstrom} to
account for both the IR and UV resonances. We show results for $T=0$
which we expect to hold even at room temperature for separations in
the range  $L<1\mu$m. We have confirmed that our 
numerical results, performed at $T=0$, are in agreement with those of 
Ref. \cite{13} in  the limit of very thick plates. 

As expected, Fig. \ref{iantfvsdvsL} shows that for very thin films 
the torque becomes null, while for thick enough films it attains its
asymptotic value corresponding to semi-infinite plates. 
Fig. \ref{iantfvsdvsL} displays small maxima corresponding to
an optimal thickness for which the torque is maximized for each value
of the separation, analogous to those found above for anisotropic
conducting films, although these maxima are very broad.
In this case, the maximum torque is about one order of magnitude smaller than
for the case of anisotropic conductors shown in
Fig. \ref{tvsdwpe_wpo2} when typical metallic plasma frequencies are
chosen, and about two orders of magnitude smaller than for the case of
the ideal uniaxial plates shown in Fig. \ref{tvsaideal}.

In order to observe the Casimir torque, several experimental setups
have been suggested \cite{13,munday2006}. If the vacuum Casimir cavity
is replaced by a cavity filled with a dielectric liquid, and if the
dielectric function of the fluid is intermediate between that of the
cavity walls, then the sign of the Casimir force may be reversed,
becoming repulsive instead of attractive. Munday et al. \cite{13} took
advantage of this sign reversal and estimated that a quartz or calcite
anisotropic disk with a diameter and a width of some tens of
$\mu m$ would float at a height of about one hundred $nm$
over the surface of a barium titanate anisotropic crystal if immersed in
ethanol. At this height, the repulsive Casimir force would balance the
weight of the disk, which would then be able to rotate freely around
its axis. Then it would be possible to first align the disk with the
polarization of a laser beam and optically monitor the rotation
towards the equilibrium orientation after the laser beam is turned
off. The characteriztic time would depend of the viscosity of the
fluid and on the magnitude of the driving Casimir torque which would
thus be obtained. 

In a second proposal by the same group \cite{munday2006} a large set of much
smaller disks of $\mu m$ scale width and diameter are kept separated a few
nanometers from the substrate not by a repulsive Casimir force but by
surfactant molecules that attach to the surfaces of both the substrate
and the disks which in this case would be immersed in an
electrolyte. The orientation of such small disks would be disrupted by
their rotational Brownian motion, but the distribution of their
orientations would yield **** Sepasuchi.

According to our calculations above, structured materials made up of
aligned nanowires embedded in dielectric matrices are subject to much
larger torques than the birefringent crystals considered my Munday et
al. \cite{13,munday2006}. Thus, the Casimir torque could be measured
using a torsion balance. To avoid alignment problems, a spherical
anisotropic particle could be used instead of a flat disk. Hanging a
50 $\mu m$ sphere from a 10 $\mu m$ long aluminum wire of 50 $nm$
radius

frecuencia =100Hz
vacío
placa giratoria. 50HZ.

\begin{comment}
Iannuzzi. Discos flotando. 

realización: nanoalambres en matriz

configuración. Esfera. 
\end{comment}
\section{Conclusions}\label{conclusion}
We have extended a fictitious-cavity approach to calculate the
Casimir force, energy and torque between anisotropic media with
a planar geometry in terms only of their optical coefficients.  
Our results are applicable to arbitrary anisotropic
materials which may be conducting or insulating, opaque or
transparent, semi-infinite or with a finite thickness, with a non-dispersive
or a frequency dependent response, homogeneous or structured, as long
as we can calculate their corresponding reflection coefficients.
Our expressions for the torque were simply obtained by analytically
deriving the energy with respect to the relative orientation
of the plates. The formalism has allowed us to perform
calculations at zero and at finite temperature.
  
We have reproduced some known results and we have calculated the
torque for ideal uniaxial systems which are the anisotropic counterparts   
to the ideal Casimir  mirrors, namely, systems which are perfect
conductors along some directions and perfect insulators along
others. We also calculated the torque for systems with a Drude conductivity.
As a non trivial application of our formalism, 
we have calculated the torque for a system consisting of two
anisotropic slabs, one of which is a semi-infinite substrate and
the other a thin film of thickness $d$. The slabs were modelled as Drude
metals and  
made up of the same material in order to study the effect of the film
thickness in the torque. Our numerical results show that exists an
optimal value of the film thickness where the torque is maximized. We
have estimated the optimal thickness in the retarded regime
by studying the anisotropy of the skin depth evaluated at the
characteristic frequency of the cavity; the maximum torque is
achieved for films whose thickness is intermediate between the largest
and the smallest of the two principal skin depths. 
Finally, we calculated the Casimir torque for a thin film of calcite
above a barium titanate substrate. The results seem qualitatively
similar although the torque is smaller than that for uniaxial
conductors and for the ideal system.
 
In conclusion, we developed a very general formalism to calculate the
Casimir torque and illustrated its use by studying the torque at zero
and finite temperatures for ideal and realistic, conducting
and insulating systems of semi-infinite and finite widths. We expect
that the simplicity and generality of our formalism motivates further
studies of these relatively unexplored aspects of the Casimir effect
between anisotropic media. 

\acknowledgments
%We acknowledge useful discussions with C. Villarreal and R. Esquivel.
This work was partially supported by DGAPA-UNAM under grant
IN120909.

\appendix
\section*{Appendix}

In this appendix we derive the expressions for the matrices  $\mathbf
t_{20}$, $\mathbf   
r_{20}$, $\mathbf
t_{02}$ and  the $z$-component of the wave-vectors
of the ordinary and extraordinary waves. We derive these expressions
by following the method 
developed in Refs. \cite{azzam,berreman}.  
We assume that the system is described by the geometry shown in 
 Fig. \ref{config}.
Thus, the dielectric tensor is given by 
\begin{equation}\label{epsrotated}
\bm\epsilon=\left[\begin{array}{ccc}
\epsilon_{\parallel}\cos^2\psi +  \epsilon_{\perp}\sin^2\psi&
  (\epsilon_{\perp} - \epsilon_{\parallel})\sin\psi\cos\psi&0\\
(\epsilon_{\perp}-\epsilon_{\parallel}) \sin\psi\cos\psi&
 \epsilon_{\perp}\cos^2\psi+ \epsilon_{\parallel}\sin^2\psi&0\\
 0&0&\epsilon_{\perp}
\end{array}\right].
\end{equation}
All of the quantities above (the dielectric functions and the angles)
refer to plate 2, but we omit the corresponding index (2) in order to
simplify our notation.
On the other hand, from Maxwell equations, the components of
the electromagnetic 
field parallel to the $x-y$ plane, can be cast in a set of four
differential equations 
\begin{equation}\label{difPsi}
\partial_z \bm F=iq \mathbf\Gamma \bm F,
\end{equation}
where 
\begin{equation}
\bm F=(E_x,B_y,E_y,-B_x)^T,
\end{equation}
$E_x$, $E_y$, $B_x$ and $B_y$  are the  electric
 and magnetic field components parallel to the interfaces, $q$ is the
 free-space wavenumber and 
$\mathbf\Gamma$ is a $4\times 4$ 
matrix with elements      
\begin{equation}\label{Gamma}
\begin{array}{ccl}
\Gamma_{11}&=&-\sin\theta\epsilon_{zx}/\epsilon_{zz},\\
\Gamma_{12}&=&1-\sin^2\theta/\epsilon_{zz},\\
\Gamma_{13}&=&-\sin\theta\epsilon_{zy}/\epsilon_{zz},\\
\Gamma_{14}&=&\Gamma_{24}=\Gamma_{31}=\Gamma_{32}=\Gamma_{33}=\Gamma_{44}=
0,\\
\Gamma_{21}&=&\epsilon_{xx}-
\epsilon_{xz}\epsilon_{zx}/\epsilon_{zz},\\
\Gamma_{22}&=&-\sin\theta\epsilon_{xz}/\epsilon_{zz},\\
\Gamma_{23}&=& \epsilon_{xy} -
\epsilon_{xz}\epsilon_{zy}/\epsilon_{zz},\\
\Gamma_{34}&=&1,\\
\Gamma_{41}&=&\epsilon_{yx}-\epsilon_{yz}\epsilon_{zx}/\epsilon_{zz},\\
\Gamma_{42}&=&\sin\theta\epsilon_{yz}/\epsilon_{zz},\\
\Gamma_{43}&=&\epsilon_{yy}-
\sin^2\theta-\epsilon_{yz}\epsilon_{zy}/\epsilon_{zz}.
\end{array}
\end{equation}
where $\theta$ is the angle of   
incidence.
Within a homogeneous medium  $\mathbf\Gamma$ is independent of $z$ and
Eq. (\ref{difPsi}) 
has four particular solutions of the form 
\begin{equation}\label{solPsi}
\bm F=\bm F_\ell(0)e^{ik_\ell z}, \quad \ell=1,2,3,4,
\end{equation}
where  $k_\ell$ is the component of the propagation vector
parallel to the $z$-axis. Substitution of Eq. (\ref{solPsi}) into the
Eq. (\ref{difPsi}) yields the eigenvalue equation 
\begin{equation}\label{eigenvalor}
(k_\ell\mathbf{I}-q \mathbf{\Gamma}) \bm F_\ell(0)=0
\end{equation}
whose eigenvalues 
\begin{equation}\label{eigenvalues}
k_\ell = \pm k_o,\pm k_e
\end{equation}
are
\begin{equation}
k_o=Jq,
\end{equation}
\begin{equation}
k_e=Iq/\sqrt{\epsilon_{\perp}}, 
\end{equation}
where  the quantities $I$ and $J$ are defined by
Eqs. (\ref{ij}). 
 The corresponding four eigenvectors are
\begin{equation}\label{eigenvectors}
\begin{array}{cc}
\bm V^o_\pm =
\left(\begin{array}{c}\pm\tan\psi/J\\ \epsilon_\perp\tan\psi/J^2
  \\\pm1/J\\1\end{array}\right),&   
\bm V^e_\pm=\left(\begin{array}{c}
  \mp\frac{J^2\cot\psi}{I\sqrt\epsilon_\perp} \\- \cot\psi \\
\pm\sqrt\epsilon_\perp/I\\ 1\end{array}\right),
\end{array}
\end{equation}
where the eigenvector $\bm V^\mu_\sigma$ describes the
$\mu$-polarized wave with a wave vector component $\sigma k_\mu$ along the
$z$-axis.

Consider the reflection of light impinging at
the interface between vacuum and a semi-infinite anisotropic medium
from within the medium. The electromagnetic 
field at the vacuum side is made up of only a transmitted 
wave. Inside the medium, the electromagnetic field is made up of four
waves $\bm F=\sum_{\sigma,\mu} \psi_\sigma^\mu \bm V_\sigma^\mu e^{i\sigma
  k_\mu z}$.  The field $\bm F$ is continuous across the interface, thus 
\begin{equation}\label{bc1}
\bm F_t= \sum_{\sigma,\mu} \psi_\sigma^\mu \bm V_\sigma^\mu e^{i\sigma
  k_\mu z}.
\end{equation}
Writing the
generalized vector field $\bm F_t$ in terms 
of the spinor $\bm\xi_t=(\xi^p_t,\xi^s_t)^T$, defined
through Eqs. (\ref{ey})-(\ref{by}), when they are applied to the
transmitted field,
 we get 
\begin{equation}\label{Ft}
\bm F_t=(-\xi_t^p\cos\theta, \xi_t^p, \xi_t^s,
-\xi_t^s\cos\theta)^T. 
\end{equation}
Substitution of this equation into Eq. (\ref{bc1}) 
yields four linear algebraic equations that can be recast as 
two matrix equations for the spinors $\bm
\xi_t$ and $\bm
\psi_\pm=(\psi_\pm^o,\psi_\pm^e)^T$,
 \begin{equation}\label{eqmatriz3}
-\cos\theta \bm \xi_t =\mathbf s_1\bm \psi_++\mathbf s_3 \bm \psi_-
\end{equation}
and
\begin{equation}\label{eqmatriz4}
\bm \xi_t =\mathbf s_2\bm \psi_++\mathbf s_4 \bm \psi_-,
\end{equation}
where $\mathbf s_1$ and $\mathbf s_2$ are the $2\times 2$-matrices 
\begin{equation}\label{s1s2}
\begin{array}{cc}
\mathbf s_1=\left(\begin{array}{cc}V^o_{1,+}& V^e_{1,+}\\ V^o_{4,+}&
V^e_{4,+}\end{array}\right),& 
\mathbf s_2=\left(\begin{array}{cc}V^o_{2,+}& V^e_{2,+}\\ V^o_{3,+}&
V^e_{3,+}\end{array}\right),
\end{array}
\end{equation}
here, $V^\mu_{i,\sigma}$ is the $i$-th component of the eigenvector
$V^\mu_\sigma$ given by Eqs. (\ref{eigenvectors}) and when these
are substituted in   
Eqs. (\ref{s1s2}) yield  the same expressions  than those of Eqs.
(\ref{s1s2r}). Similarly, $\mathbf s_3$ and $\mathbf s_4$ are
the matrices  
\begin{equation}\label{s3s4}
\begin{array}{cc}
\mathbf s_3=\left(\begin{array}{cc}V^o_{1,-}& V^e_{1,-}\\ V^o_{4,-}&
V^e_{4,-}\end{array}\right),& 
\mathbf s_4=\left(\begin{array}{cc}V^o_{2,-}& V^e_{2,-}\\ V^o_{3,-}&
V^e_{3,-}\end{array}\right).
\end{array}
\end{equation}
 Elimination of $\bm\xi_t$ from
 Eqs. (\ref{eqmatriz3})-(\ref{eqmatriz4}) yields   
\begin{equation}\label{prer20}
\bm \psi_+= -(\mathbf s_1 +\mathbf s_2\cos\theta
)^{-1}(\mathbf s_3 
+\mathbf s_4\cos\theta) \bm \psi_-.
\end{equation}
Thus, from $\bm \psi_+=\mathbf r_{20} \bm \psi_-$ we obtain
\begin{equation}\label{r20}
\mathbf r_{20}= -(\mathbf s_1 +\mathbf s_2\cos\theta
)^{-1}(\mathbf s_3 
+\mathbf s_4\cos\theta). 
\end{equation}
The matrix $\mathbf t_{20}$ can be obtained by substituting
Eq. (\ref{prer20}) into Eq. (\ref{eqmatriz4}) to get
\begin{equation}
\bm\xi_t=(\mathbf s_4 +\mathbf s_2
\mathbf r_{20}) \bm \psi_-, 
\end{equation}
from which we obtain 
\begin{equation}\label{t20}
\mathbf t_{20} =(\mathbf s_4 +\mathbf s_2
\mathbf r_{20}). 
\end{equation}
Finally, in order to calculate the matrices $\mathbf t_{02}$ and
$\mathbf r_{02}$, we consider the reflection of light impinging at
a semi-infinite anisotropic medium
from vacuum. The electromagnetic 
field at the medium side is made up of two transmitted 
waves, $\bm F=\sum_{\mu} \psi_+^\mu \bm V_+^\mu e^{i k_\mu
  z}$. Imposing continuity on $\bm F$ across the interface  
\begin{equation}\label{bc2}
\bm F_i+ \bm F_r= \sum_{\mu} \psi_+^\mu \bm V_+^\mu e^{ik_\mu z}.
\end{equation}
Writing the generalized vector field $\bm F_i$ and $\bm F_r$ in terms 
of the spinors $\bm\xi_i=(\xi^p_i,\xi^s_i)^T$ and
$\bm\xi_r=(\xi^p_r,\xi^s_r)^T$  we get 
\begin{equation}\label{Fi}
\bm F_i=(\xi_i^p\cos\theta, \xi_i^p, \xi_i^s,
\xi_i^s\cos\theta)^T 
\end{equation}
and
\begin{equation}\label{Fr}
\bm F_r=(-\xi_r^p\cos\theta, \xi_r^p, \xi_r^s,
-\xi_r^s\cos\theta)^T. 
\end{equation}
Substitution of this equation into Eq. (\ref{bc2}) 
yields four linear algebraic equations that can be recast as 
two matrix equations for the spinors $\bm \psi_+$, $\bm\xi_i$ and $\bm\xi_r$
\begin{equation}\label{eqmatriz5}
\cos\theta(\bm \xi_i-\bm\xi_r)=\mathbf s_1\bm \psi_+
\end{equation}
and
\begin{equation}\label{eqmatriz6}
\bm \xi_i+\bm\xi_r =\mathbf s_2\bm \psi_+.
\end{equation}
Elimination of $\bm\psi_+$ from the last two equations yields
\begin{equation}\label{r02}
\bm \xi_r= \mathbf r_{02} \bm \xi_i, 
\end{equation}
where $\mathbf r_{02}$ is given by Eq. (\ref{r}).
Elimination of $\bm \xi_r$ from
Eqs. (\ref{eqmatriz5})-(\ref{eqmatriz6}) yields 
\begin{equation}\label{t02}
\mathbf t_{02}=\mathbf s_1^{-1}\cos\theta(\mathbf I -\mathbf r_{02}).
\end{equation}
This concludes the calculation of the reflection and transmission
matrices corresponding to the front surface of the film.

\end{document}